\documentclass[12pt]{article}
\usepackage{epsfig}  
\begin{document}
\suppressfloats
\title{
Orbital-free effective embedding potential at nuclear cusp
}
\author{Juan Maria Garcia Lastra$^a$, Jakub W. Kaminski, \\ and Tomasz A. Weso{\l}owski\\
  Universit\'e de Gen\`eve, \\
                D\'epartement de Chimie Physique\\
                30, quai Ernest-Ansermet, \\
                CH-1211 Gen\`eve 4, Switzerland \\
  $^a$ Universidad del Pa\'{\i}s Vasco\\
  Departamento de F\'{\i}sica de Materiales\\
 E-20018 Donostia-San Sebastian, Spain 
 }
\date{\today}
\maketitle
\begin{abstract} 
A new approach to approximate the kinetic-energy-functional dependent
component ($v_t[\rho_A,\rho_B](\vec{r})$) of
the effective potential in one-electron equations for orbitals embedded in a frozen density
environment  (Eqs. 20-21 in [Wesolowski and Warshel, {\it J. Phys. Chem.} {\bf 97},
(1993) 8050]) is proposed.
The exact limit for $v_t$ at
$\rho_A\longrightarrow 0$ and $\int \rho_B d\vec{r}=2$ is enforced.
The significance of this limit is analysed formally and numerically
for model systems including a numerically solvable model and real
cases where $\int \rho_B d\vec{r}=2$.
A simple approximation to $v_t[\rho_A,\rho_B](\vec{r})$ is constructed
which enforces the considered limit near nuclei in the environment.
Numerical examples are provided to illustrate the numerical significance
of the considered limit for real systems -
intermolecular complexes comprising, non-polar, polar, charged
constituents.
Imposing the limit  improves significantly the quality of the approximation to
$v_t[\rho_A,\rho_B](\vec{r})$ for systems comprising charged
components. For complexes comprising neutral molecules or atoms the
improvement occurs as well but it is numerically insignificant.
\end{abstract}

\section{Introduction}

Numerical methods to study electronic structure in condensed matter
use mainly  techniques developed for periodic systems.%
In many cases, however, methods developed for finite systems
are also used. They are especially adequate for ionic solids, liquids,
molecular crystals, clusters of molecules, for instance, to study  features
of the electronic structure which are local in character.
In such a case, the electronic structure is modelled only in some
well-defined region in space of direct relevance. The effect of the atoms outside of this  selected
region (referred to as environment in this work)
is taken into account by some {\it embedding
  potential}. 
Different strategies are  applied in practice to represent the embedding
potential. They differ in the choice of descriptor of the
environment.
The roughest approximation is to neglect the environment entirely.
Such simplification is commonly used to study chemical bonding and
reactivity in condensed phase if the solvent in which the reaction takes
place is known to play a secondary role. 
Representing the environment (discrete or continuous,
polarisable or not) by the electric field it generates, makes it
possible to take into account the effect of the environment~\cite{PCMsolv,tapia1975}.
Such 
classical treatment of the effect of the environment on the electronic
structure
is commonly used both in chemistry and in materials science~(for
review see Ref.~\cite{QMMMrev}).
The embedding potential in such methods is obviously orbital-free.
It is, however, not exact because the quantum statistics nature of
electrons is completely neglected.
Taking into account the fermion nature of electrons might proceed by
following a similar strategy as the one applied by Phillips and
Kleinman
in the construction of pseudopotentials in order to eliminate explicit
treatment of core electrons~\cite{phillipskleinman}.
For recent developments along these lines, see
Ref.~\cite{Henderson2006}.

Using the following elements of the Hohenberg-Kohn-Sham formulation of density
functional theory: Hohenberg-Kohn theorems~\cite{HK1964}, 
a reference system of non-interacting electrons~\cite{KS1965},
and the corresponding density functional of the kinetic energy
($T_s[\rho]$)~\cite{Levy_CS}
in particular, leads to the embedding potential which is {\it exact}
in the limit of exact functionals and {\it orbital-free} i.e. does not
involve other descriptors of the environment than its electron density~\cite{WW1993}.
The pure-state non-interacting $v$-representable electron density
$\rho_A^{min}$, 
such that added to some arbitrarily
chosen density associated with the environment ($\rho_B$) minimises the
Hohenberg-Kohn energy functional for the whole system, 
can be obtained from the following one-electron
equations  (Eqs. 20-21 in Ref.~\cite{WW1993}):
\begin{eqnarray}
\left[-\frac{1}{2}\nabla^2+v_{eff}^{KSCED}\left[\rho_A,\rho_B;\vec{r}\right]\right]\phi^A_{i}=\epsilon^A_{i}\phi^A_{i}
 \;\;\;\;\; i=1,N^{A}  \label{ksceda}
 \\ \nonumber
\end{eqnarray}
where $\rho_A= 2\sum_i^{N_A} \left\vert \phi_i\right\vert ^2$ and
and $v_{eff}^{KSCED}\left[\rho_A,\rho_B;\vec{r}\right]$
denotes a local potential which depends only on electron densities
$\rho_A$ and $\rho_B$.
The label KSCED  (Kohn-Sham Equations with Constrained Electron Density)
is used here to indicate that the  
local potential differs from that in Kohn-Sham equations~\cite{KS1965} for either the total
system ($v^{KS}[\rho_A+\rho_B;\vec{r}])$) or 
the isolated subsystem $A$
($v^{KS}[\rho_A;\vec{r}])$).
Also the one-electron functions
($\lbrace\phi^A_{i}\rbrace$) obtained from Eq.~\ref{ksceda} are not optimal orbitals in neither Kohn-Sham systems.
Atomic units are applied in all formulas which are given for spin
unpolarised systems.

The total effective potential in Eq.~\ref{ksceda} is the sum of
  the
conventional Kohn-Sham effective potential 
$v_{eff}^{KS}\left[\rho_A+\rho_B;\vec{r}\right]$
for the whole  system
evaluated for the electron density $\rho=\rho_A+\rho_B$
and another local potential ($v_t[\rho_A,\rho_B](\vec{r})$):
\begin{eqnarray}
v_{eff}^{KSCED}\left[\rho_A,\rho_B;\vec{r}\right]=
v_{eff}^{KS}\left[\rho_A+\rho_B;\vec{r}\right]+ v_t[\rho_A,\rho_B](\vec{r})                                                      
\label{kscedpot0}
 \\ \nonumber                           
\end{eqnarray}
where  $v_t[\rho_A,\rho_B](\vec{r})$ involves functional derivatives
of the functional $T_s[\rho]$:
\begin{eqnarray}
v_t(\vec{r})=v_t[\rho_A,\rho_B](\vec{r})=
\left.\frac{\delta T_{s}[\rho]}{\delta
    \rho}\right|_{\rho=\rho_A+\rho_B} 
-
\left.\frac{\delta T_{s}[\rho]}{\delta
    \rho}\right|_{\rho=\rho_A} 
 \label{definition_of_vt} 
\end{eqnarray}
Note that no restriction is made concerning the overlap between $\rho_A$ and
$\rho_B$ in real space.
The potential  $v_t[\rho_A,\rho_B](\vec{r})$
can be alternatively expressed as:
\begin{eqnarray}
v_t[\rho_A,\rho_B](\vec{r})=
\left.\frac{\delta T_{s}^{nad}[\rho,\rho_{B}]}{\delta
    \rho}\right|_{\rho=\rho_A}  \label{definition_of_vt_alt} 
\end{eqnarray}
where $T_s^{nad}[\rho_A,\rho_B]$ denotes the following difference:
\begin{eqnarray}
T_s^{nad}[\rho_A,\rho_B]=T_s[\rho_A+\rho_B]-T_s[\rho_A]-T_s[\rho_B]\label{tsnad_def}
 \\ \nonumber
\end{eqnarray}

For the sake of the subsequent discussions, it is convenient to split 
the total effective potential $v_{eff}^{KSCED}\left[\rho_A,\rho_B;\vec{r}\right]$
into two components: the Kohn-Sham effective potential for the
isolated subsystem $A$
($v_{eff}^{KS}\left[\rho_A;\vec{r}\right]$),
which is $\rho_B$ independent, and 
the remaining part representing the environment:\\
\begin{eqnarray}
v_{eff}^{KSCED}\left[\rho_A,\rho_B;\vec{r}\right]=
v_{eff}^{KS}\left[\rho_A;\vec{r}\right]+                                                      
v_{emb}^{KSCED}\left[\rho_A,\rho_B;\vec{r}\right] \label{kscedpot1}
 \\ \nonumber                           
\end{eqnarray}
where
\begin{eqnarray}
v_{emb}^{KSCED}[\rho_{A},\rho_{B};\vec{r}]&=&
v_{ext}^B(\vec{r}) + 
\int \frac{\rho_{B}(\vec{r}')}{|\vec{r}'-\vec{r}|} d\vec{r}'
\label{kscedembpot}
\\
 \nonumber \\
&+&\left.\frac{\delta E_{xc}\left[\rho\right]}{\delta\rho}\right\vert_{\rho=\rho_A+\rho_B}-
\left.\frac{\delta
    E_{xc}\left[\rho\right]}{\delta\rho}\right\vert_{\rho=\rho_A}+
v_t[\rho_A,\rho_B](\vec{r})
\nonumber \\ \nonumber
\end{eqnarray}
where $E_{xc}[\rho]$ denotes the Kohn-Sham functional of the
exchange-correlation energy~\cite{KS1965}.

Orbital-free effective embedding potential given in
Eq.~\ref{kscedembpot},
and its $v_t[\rho_A,\rho_B](\vec{r})$ component in particula, are used
in various types of multi-level numerical simulations (for a review, see Ref.~\cite{wesolowski_WS}
or
Refs.~\cite{truong,MeiPRB00,TrailBird2000,Zbiri2004,Nakano2005,Neugebauer_excitations2,warshel.applic2,Choly2005,jacobNMR,neugebauercd2006}
for representative recent reports).
Such simulations deal with condensed matter systems, for which the
electronic features of a selected subsystem (subsystem $A$) are subject to detailed
investigation whereas $\rho_B$ is subject to additional
simplifications.
Other formal frameworks use also  $v_t[\rho_A,\rho_B](\vec{r})$ such as: 
  Cortona's formulation of density functional
theory~\cite{Cortona1991},  where $\rho_B$ is not an assumed quantity 
but a result of fully variational calculations~\cite{Cortona1991,WW1996,hutter,WesolowskiTran2003,Dulak2007}
or linear-response time-dependent density-functional-theory
description of electronic excitations localised in embedded
systems~\cite{Casidawesolowski,WesolowskiJACS2004}.
Finally, the orbital-free effective embedding potential given in
Eq.~\ref{kscedembpot},
and its $v_t[\rho_A,\rho_B](\vec{r})$ component in particular,
are used in combination with traditional wave-function based methods by Carter
and collaborators
(see for instance Ref.~\cite{K2002}).
For the formal analysis of applicability of 
such a combination, see Ref.~\cite{Wesolowski:PRA:77:012504:2008}
which shows that the exact embedding potential in such a case 
always comprises the $v_t[\rho_A,\rho_B](\vec{r})$ component.

In practical applications,  $v_t[\rho_A,\rho_B](\vec{r})$ is not used
but some analytic expressions approximating this quantity 
($\tilde{v}_t[\rho_A,\rho_B](\vec{r})$) for the obvious sake of
practical advantages.
This replacement results in errors in all derived quantities
which will be referred to as {\it $\tilde{v}_t$-induced errors}.
In each case,  $\tilde{v}_t[\rho_A,\rho_B](\vec{r})$
is obtained by using analytic form of an approximated functional
$\tilde{T}_s[\rho]$ into Eqs.~\ref{definition_of_vt}-\ref{tsnad_def}.
We will refer to such
$\tilde{v}_t[\rho_A,\rho_B](\vec{r})$
as {\it decomposable} because 
the analytic form of all relevant quantities: $\tilde{T}_s[\rho]$,
$\tilde{T}_s^{nad}[\rho_A,\rho_B]$,
and $\tilde{v}_t[\rho_A,\rho_B](\vec{r})$, is available.
If  the form of the used
$\tilde{T}_s[\rho]$ comprises only low-level  gradient-expansion~\cite{Kirzhnitz}
contributions, the corresponding decomposable  $\tilde{v}_t[\rho_A,\rho_B](\vec{r})$
violates the exact limit for ${v}_t[\rho_A,\rho_B](\vec{r})$ 
at $\rho_A\longrightarrow 0$ and $\int \rho_B d\vec{r}=2$ (see
Appendix A).
Our interest in  the local behaviour of
${v}_t[\rho_A,\rho_B](\vec{r})$ at this limit is motivated by the
fact that the corresponding conditions%
occur  if
$\rho_B$ comprises two electrons tightly bound to a distant
nucleus in the environment such as in the case of
the helium atom, Li$^+$ cation,  Be$^{2+}$, etc.
We expect that they are relevant also for heavier nuclei if in a
volume element centred on the nucleus $\rho_B$ is dominated by a
doubly-occupied orbital.

The present work focuses on the investigation whether the considered
exact limit is of any  practical relevance.
To this end, we apply the following strategy:
{\it i}) We use a model system (Appendix B), for which the conditions
 $\rho_A\longrightarrow 0$ and
$\int \rho_B d\vec{r}=2$
apply rigorously, to analyse the importance of enforcing the
correct local behaviour of  $\tilde{v}_t[\rho_A,\rho_B](\vec{r})$.
{\it ii}) We construct a simple approximation to
${v}_t[\rho_A,\rho_B](\vec{r})$
obeying the considered exact limit in the vicinity of nuclei and
analyse the numerical significance of imposing the considered
condition in real systems where the conditions  $\rho_A\longrightarrow 0$ and
$\int \rho_B d\vec{r}=2$ do not apply rigorously.

Our ultimate goal is a new approximation to
$v_t[\rho_A,\rho_B](\vec{r})$
which can be inexpensively evaluated in practice and obeys as much as
possible of the relevant exact properties.
It should pointed out in this context that the  position-dependency of
$v_t[\rho_A,\rho_B](\vec{r})$ is the result of non-homogeneity of
$\rho_A$ and/or $\rho_B$. Therefore, the symbol
$v_t[\rho_A,\rho_B](\vec{r})$  (or
$\tilde{v}_t[\rho_A,\rho_B](\vec{r})$ if approximated)
is used throughout this work to indicate that this local quantity 
is a functional of $\rho_A$ and $\rho_B$.
Explicit position dependence is strongly undesired in
density-functional-theory based methods because it is not
straightforward to obtain 
{\it i}) such potential as a functional derivative of some density
functional, and {\it ii})  functional derivatives of such potential needed in
some formal frameworks~\cite{Casidawesolowski,WesolowskiJACS2004}.
General symbols  such $v_t$ or $v_t(\vec{r})$ are used in some
discussions where the issue of explicit position-dependence is not
relevant.

\section{Conventional (decomposable) strategies to approximate  $v_t[\rho_A,\rho_B](\vec{r})$}

Before proceeding to the construction of the desired approximation to
$v_t[\rho_A,\rho_B](\vec{r})$ obeying the considered limit, 
we overview the conventional construction of approximation to
$v_t[\rho_A,\rho_B](\vec{r})$ and the local behaviour near a nucleus 
of the obtained
potential.
The conventional strategy, which is applied in our own works and the works by
others so far, is  to start from some explicit
density functional $\tilde{T}_s[\rho]$ and to use its analytic form to derive
the corresponding approximate expression for $T_s^{nad}[\rho_A,\rho_B]$:
\begin{eqnarray}
T_s^{nad}[\rho_A,\rho_B]\approx \tilde{T}_s^{nad}[\rho_A,\rho_B]= 
\tilde{T}_s[\rho_A+\rho_B]- \tilde{T}_s[\rho_A] -
\tilde{T}_s[\rho_B]\label{decomposableGE}
\end{eqnarray}
and to use the obtained analytic expression to
 obtain 
$\tilde{v}_t[\rho_A,\rho_B](\vec{r})$  by means of functional differentiation.
\begin{eqnarray}
v_t[\rho_A,\rho_B](\vec{r})\approx \tilde{v}_t[\rho_A,\rho_B](\vec{r}) =
\left.\frac{\delta \tilde{T}_{s}^{nad}[\rho,\rho_{B}]}{\delta
    \rho}\right|_{\rho=\rho_A}  \label{decomposablev} 
\end{eqnarray}
This strategy can be applied for any approximated functional
$\tilde{T}_s[\rho]$ provided its form
makes it possible to obtain the 
analytic expression for
$\tilde{v}_t[\rho_A,\rho_B](\vec{r})$.
Simple functionals  $\tilde{T}_s[\rho]$, which
depend explicitly on densities and their gradients, are of particular
practical interest.
They lead to
$\tilde{v}_t[\rho_A,\rho_B](\vec{r})$
which
depends explicitly only on $\rho_A$ and $\rho_B$
and their first- and second derivatives.
In the original work by Cortona~\cite{Cortona1991},
where the subsystem formulation of density functional theory  was introduced,
a decomposable
$\tilde{v}_t[\rho_A,\rho_B](\vec{r})$
derived from the 
Thomas-Fermi~\cite{ThomasFermi}
kinetic energy functional
was used  to study ionic solids.
In our own works, only decomposable
$\tilde{v}_t[\rho_A,\rho_B](\vec{r})$ derived from gradient-dependent
approximations to
$T_s[\rho]$ were considered so far
(see for instance the analyses of their accuracy 
in Ref.~\cite{Wesolowski1997a,bernard2008} or their recent applications 
in multi-level computer simulations of condensed
matter~\cite{Neugebauer_excitations2}).

Thomas-Fermi kinetic energy
functional~\cite{ThomasFermi}, which is exact for the uniform electron
gas, leads to the following approximate expression for $T_{s}^{nad}[\rho_{A},\rho_{B}]$:
\begin{eqnarray}
\tilde{T}_{s}^{nad(TF)}[\rho_{A},\rho_{B}]=
C_{TF}\int \left(
\left(\rho_A+\rho_B\right)^{5/3}
-\rho_A^{5/3}
-\rho_B^{5/3}
\right)d\vec{r} \label{tsnad_gea0}
\end{eqnarray}
where
$C_{TF}=\frac{3}{10}(3\pi^2)^{2/3}$.\\
The associated expression for  
$\tilde{v}_t^{TF}[\rho_A,\rho_B](\vec{r})$
reads:
\begin{eqnarray}
\tilde{v}_t^{TF}[\rho_A,\rho_B](\vec{r})=
\frac{5}{3}C_{TF}\left(
\left(\rho_A+\rho_B\right)^{2/3}
-\rho_A^{2/3}
\right) \label{vtsnad_gea0}
\end{eqnarray}

Approximating $T_s[\rho]$ by means of the gradient expansion of the kinetic
 energy~\cite{Kirzhnitz} truncated to the second order
leads to the following approximate expression for $T_{s}^{nad}[\rho_{A},\rho_{B}]$~\cite{WW1993}:
\begin{eqnarray}
\tilde{T}_{s}^{nad(GEA2)}[\rho_{A},\rho_{B}]=T_{s}^{nad(TF)}[\rho_{A},\rho_{B}]-
\frac{1}{72}
\int\frac{|\rho_A\nabla\rho_B-\rho_B\nabla\rho_A|^2}{\rho_A\rho_{B}(\rho_A+\rho_B)}d\vec{r}
\label{tsnad_gea2} \\ \nonumber
\end{eqnarray}
The associated expression for  
$\tilde{v}_t^{GEA2}[\rho_A,\rho_B](\vec{r})$
is given in Ref.~\cite{WW1993}.

For the the group of gradient-dependent approximations to $T_s[\rho]$
of the {\it generalised
   gradient approximation} form~\cite{Wesolowski1997a,Wesolowski1996}
the analytic expression  for $\tilde{T}_{s}^{nad}[\rho_{A},\rho_{B}]$
reads:
\begin{eqnarray}
\tilde{T}_{s}^{nad(GGA)}[\rho_{A},\rho_{B}]=C_{TF}\int \left[ 
\left(\rho_A+\rho_B\right)^{5/3} F(s_{AB})
-\rho_A^{5/3} F(s_{A})
-\rho_B^{5/3} F(s_{B})
\right] d\vec{r}
\label{tsnad_gga} \\ \nonumber
\end{eqnarray}
where $F^{GGA}(s)$ ({\it enhancement
  factor}) depends on a  dimensionless quantity
$s=\frac{\vert\nabla\rho\vert}{2(3\pi^2)^{1/3}\rho^{4/3}}$ ({\it
  reduced density gradient}). Various analytic forms of  $F^{GGA}(s)$
were
  proposed
in the literature~\cite{LLP,LC94,Fuentealba,Tran2002a}.
The associated analytic expression for 
$\tilde{v}_t^{GGA}[\rho_A,\rho_B](\vec{r})$
is given in Ref.~\cite{WesolowskiTran2003}.
It is worthwhile to notice that the GGA form is flexible and includes
$\tilde{T}^{TF}[\rho]$
and  $\tilde{T}^{GEA2}[\rho]$ as special cases.
Numerical values of $s$ provide useful information about shell
structure and the distance from the nucleus in atoms~\cite{Zupan1997}. 
For an atom, $s$ is known to be small near the nucleus, reach the
values of about 3 in the valence region, and 
diverge exponentially to $+\infty$ at large distances. 
In molecules, it behaves similarly with a noticeable exception of
stationary points of electron density (bond midpoints for instance)
where $s=0$. 
Each approximated functional given in
Eqs.~\ref{tsnad_gea0}-\ref{tsnad_gga} comprises a dominant Thomas-Fermi
component and satisfies two  exact conditions:
\begin{itemize}
\item $T_s[\rho_A+\rho_B]-T_s[\rho_A]-T_s[\rho_B]=0$ for
non-overlapping $\rho_A$ and $\rho_B$.
\item For uniform  $\rho_A$ and $\rho_B$, they recover the exact
      analytical expression for
      $T_s[\rho_A+\rho_B]-T_s[\rho_A]-T_s[\rho_B]$.
\end{itemize}

The common feature of each among the above approximations for $T_s[\rho]$  
is that none of them yields the exact analytic form of
${v}_t[\rho_A,\rho_B](\vec{r})$ 
 at $\rho_A\longrightarrow 0$
and $\int \rho_B d\vec{r}=2$ (see Appendix A): 
\begin{eqnarray}
{v}_t[\rho_A,\rho_B](\vec{r})
{\longrightarrow}
v_t^{limit}[\rho_B](\vec{r})= \frac{1}{8}\frac{\vert \nabla \rho_B\vert^2}{\rho_B^2} -
\frac{1}{4}\frac{\nabla^2\rho_B}{\rho_B}
\label{vsnadw} \\ \nonumber
\end{eqnarray}

\begin{figure}[h]             %
\centerline{\psfig{file=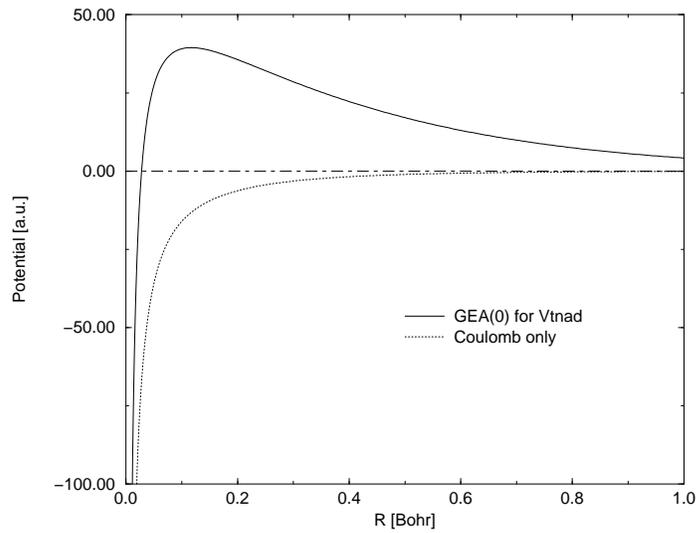,width=3.3in,angle=-90}}
\caption{
Effective potential (Eq.~\ref{kscedembpot}) calculated using local density approximation for
its exchange- and kinetic energy components for:
$\rho_B=\rho_{He}$, $\rho_A\longrightarrow 0$, and $v_{ext}^B(\vec{r})=-2/r$.
\label{LDA_embedding_potential}}                                                                                  
\end{figure}

The expression given in Eq.~\ref{vtsnad_gea0}, which
provides the dominant contribution to gradient-expansion based approximations to
$v_t[\rho_A,\rho_B](\vec{r})$ does not comprise the relevant term at
all whereas  the second-order term
provides only 1/9 of the exact expression.
Figure~\ref{LDA_embedding_potential} shows $\tilde{v}_{emb}^{KSCED(LDA)}[\rho_A,\rho_B](\vec{r})$
for  a spherically symmetric case: $\rho_B=\rho_{He}$, $v_{ext}^B(\vec{r})=-2/r$, and
$\rho_A\longrightarrow 0$, which represents a 
helium atom far from
subsystem $A$. In the Figure as well as in the following discussion, $r$ denotes the
distance from the considered nucleus. 
The potential in the figure shows the features which are
common also for heavier  atoms  if gradient expansion based
approximations to $T_s[\rho]$ are used to derive $v_t[\rho_A,\rho_B](\vec{r})$: a  very narrow
and deep  well
(reaching $-\infty$) centred on the nucleus and surrounding it
repulsive shell. 
The ${v}_t[\rho_A,\rho_B](\vec{r})$  component of the shown embedding potential is finite at the
nucleus
instead of behaving as $\frac{\zeta}{r}$.
Note that the exact term has the same form as the potential due to Coulomb
attraction by the nucleus of the charge $Z$.
Therefore, the exact $v_t$ partially compensates this attraction to
some extend because $\zeta$ is smaller than $Z$~\cite{clementiraimondi1963}.
For the particular case considered in the Figure, the missing
$\frac{\zeta}{r}$ component does not lead to any bound states
(Appendix B).
In general, however, the wrong asymptotic of the singularity at the nucleus, can lead to an unphysical transfer of electron
density from the investigated system to its environment 
(charge-leak~\cite{Dulak2006}).
This can occur if
the artificially  attractive (not sufficiently repulsive) approximation
to the orbital-free
effective embedding potential generates a bound state in the environment
of the energy which is
lower than the eigenvalue of the
highest occupied embedded orbital associated with embedded subsystem
$A$.
Numerical cases confirming such scenario are known
~\cite{jacob,dulak_private}.
Moreover, the numerical solution 
of the Schr\"odinger equation with
$\tilde{v}_{emb}^{KSCED(LDA)}[\rho_A,\rho_B](\vec{r})$ 
for  $\rho_B=\rho_{Li^+}$ and the external potential of the
$-\frac{3}{r}$ form, shows  a
deeply lying node-less bound state of the energy -0.209665~hartree with the maximum
of the radial electron density at $r^{max}=2.912$~bohr (Appendix B). 
The above observations indicate that
decomposable $\tilde{v}_t[\rho_A,\rho_B](\vec{r})$ obtained from  
low-order gradient-expansion based approximations to $T_s[\rho]$ 
might not be adequate for, at least, Li$^+$ cations in the environment.
The fact that the bound state associated with an atom in the
environment is too tightly bound to the nucleus indicate, however, that
the problem might be also present in atoms comprising more electrons.

In a subsequent section, a simple approximation to
${v}_t[\rho_A,\rho_B](\vec{r})$ is constructed based on these
observations. 
In principle, the exact limit should be applied in any
volume element in which $\rho_A$ vanishes and $\rho_B$ is obtained
from a doubly occupied orbital. In the proposed construction,
the exact limit is imposed only at volume elements near
heavier-than-hydrogen nuclei 
expecting that the considered condition is most relevant there.

\section{Building-in the exact limit for 
  $v_t[\rho_A,\rho_B](\vec{r})$ at $\rho_A\longrightarrow 0$ and $\int\rho_B
      d\vec{r}=2$.}

The aforementioned flaws of decomposable strategy to construct
gradient- and Laplacian dependent approximations to  ${v}_t[\rho_A,\rho_B](\vec{r})$
suggest a  bottom-up approach in which 
${v}_t[\rho_A,\rho_B](\vec{r})$ is directly a target.
A given approximated potential $\tilde{v}_t[\rho_A,\rho_B](\vec{r})$ will
be referred to as {\it non-decomposable}
if the analytic form of its two individual components
$\left.\frac{\delta
  \tilde{T}_s[\rho]}{\delta\rho}\right|_{\rho=\rho_A+\rho_B}$ 
and
$\left.\frac{\delta
  \tilde{T}_s[\rho]}{\delta\rho}\right|_{\rho=\rho_A}$
cannot be reconstructed.
The non-decomposable strategy is motivated by the fact
that there are exact properties of
 ${v}_t[\rho_A,\rho_B](\vec{r})$, which can be taken into
account quite easily in 
 $\tilde{v}_t[\rho_A,\rho_B](\vec{r})$, whereas
building-in them  into some approximate functional $\tilde{T}_s[\rho]$
is less straightforward. 
Abandoning the decomposable strategy is motivated also by
the results of our recent dedicated studies of the accuracy of various
gradient-dependent approximations to $T_s^{nad}[\rho_A,\rho_B]$, which
revealed that 
there is no correlation between the accuracy of
$\tilde{T}_s^{nad}[\rho_A,\rho_B]$,
$\tilde{v}_t[\rho_A,\rho_B](\vec{r})$
and the errors in the parent gradient-dependent approximation to
$\tilde{T}_s[\rho]$~\cite{bernard2008}.
It should be also pointed out that, that the individual
contributions $\tilde{T}_s^{nad}[\rho_A,\rho_B]$ are not needed in practice.

The non-decomposable approximation to ${v}_t[\rho_A,\rho_B](\vec{r})$
is constructed by enforcing the following exact conditions into its analytic
form:
\begin{itemize}
\item   $\tilde{T}_s^{nad}[\rho_A,\rho_B]\longrightarrow \tilde{T}_s^{nad(LDA)}[\rho_A,\rho_B]$ 
 for uniform $\rho_A$ and $\rho_B$.
\item $\tilde{T}_s^{nad}[\rho_A,\rho_B]\longrightarrow
      0$ 
for non-overlapping $\rho_A$ and $\rho_B$.
\item  $\tilde{v}_t[\rho_A,\rho_B]\longrightarrow v_t^{limit}[\rho_B]$ at $\rho_A\longrightarrow 0$ and $\int\rho_B
      d\vec{r}=2$.
\end{itemize}

The first two conditions%
are automatically satisfied
by the decomposable gradient-expansion based approximations discussed in the previous section.
Since such approximations proved to be sufficiently accurate for many
systems the same conditions are retained in the new construction.
The last condition is the key element of the present construction.

Before proceeding to the construction of the approximation obeying the
considered exact condition we note that
${v}_t[\rho_A,\rho_B](\vec{r})$
can be alternatively expressed as:
\begin{eqnarray}
{v}_t[\rho_A,\rho_B]= 
 \tilde{v}_t^{decomposable}[\rho_A,\rho_B]+f[\rho_A,\rho_B]\cdot
v_t^{limit}[\rho_B] \label{vtnad_alt}
\end{eqnarray}
All functionals in the above equation are determined locally and the
argument $\vec{r}$ is not written explicitly for simplicity
and the functionals ${v}_t[\rho_A,\rho_B]$ and $f[\rho_A,\rho_B]$
are simply related
($f[\rho_A,\rho_B]=\frac{v_t[\rho_A,\rho_B]-\tilde{v}_t^{decomposable}[\rho_A,\rho_B]}{v_t^{limit}[\rho_A,\rho_B]}$
if $v_t^{limit}[\rho_A,\rho_B]$ is non-zero).
The above form of ${v}_t[\rho_A,\rho_B]$
provides a  convenient for construction of approximation.
It can be used for any decomposable approximation 
to  ${v}_t[\rho_A,\rho_B]$, which violaties the considered condition, and 
the functional $f[\rho_A,\rho_B]$ has a clear physical meaning as a
{\it switching factor}  determining whether it is needed  to add
locally the missing component of the embedding potential.
As far as the choice for the decomposable component,
both the gradient-free potential given in Eq.~\ref{vtsnad_gea0}
and the decomposable potential derived from the Lembarki-Chermette~\cite{LC94}
approximation to $T_s[\rho]$ were shown in
dedicated studies~\cite{Wesolowski1997a,bernard2008}  to be reasonably
accurate
if  $\rho_A$ and $\rho_B$ do not overlap 
strongly. Both these approximations comprise the zeroth order 
contribution.
If  $\tilde{v}_t^{TF}[\rho_A,\rho_B]$ is used as decomposable
component in Eq.~\ref{vtnad_alt}, the terms $v_t^{limit}[\rho_B]$ and
 $\tilde{v}_t^{decomposable}[\rho_A,\rho_B]$ are local functions
 depending explicitely on $\rho_A$, $\rho_B$, $\nabla\rho_B$, and
 $\nabla^2\rho_B$.
Approximating  $f[\rho_A,\rho_B]$ by a local function depending
explicitely on these quantities applied in Eq.~\ref{vtnad_alt}
leads to an approximated potential requiring a similar computational effort as
conventional low-order gradient-expansion based decomposable
approximations to  ${v}_t[\rho_A,\rho_B]$.
The first approximation made here is replacing the switching factor
defined in Eq.~\ref{vtnad_alt} by
a {\it switching function}:
\begin{eqnarray}
f[\rho_A,\rho_B](\vec{r})\approx\tilde{f}(\rho_A,\rho_B,\nabla\rho_B,\nabla^2\rho_B)
\end{eqnarray}

The above considerations lead to the following general form of the
approximation to
$v_t[\rho_A,\rho_B](\vec{r})$:
\begin{eqnarray}
\tilde{v}_t[\rho_A,\rho_B]=
 \tilde{v}_t^{TF}[\rho_A,\rho_B]+\tilde{f}(\rho_A,\rho_B,\nabla\rho_B,\nabla^2\rho_B)
\cdot v_t^{limit}[\rho_B] \label{NDSD}
\end{eqnarray}
The above general form provides a clear interpretation for the
switching function, which can be used as  guideline in construction
of approximations -  it   
``detects'' such volume elements
for which the conditions $\rho_A\longrightarrow 0$ and $\int\rho_B
      d\vec{r}=2$ are most relevant.

\subsection{The switching function $\tilde{f}$
  for environments comprising one-nucleus and two-electrons}

In constructing $\tilde{f}(\rho_A,\rho_B,\nabla\rho_B,\nabla^2\rho_B)$
the following additional requirements  (simplifications) are made:
\begin{itemize}
\item $\tilde{f}(\rho_A,\rho_B,\nabla\rho_B,\nabla^2\rho_B)$
      is one in the vicinity of a nucleus to account fully for the missing $\frac{\zeta}{r}$
      component. 
\item The criterion for determining the range at which
      $v_t^{limit}[\rho_B]$ 
      is nuclear number independent.
\item $\tilde{f}(\rho_A,\rho_B,\nabla\rho_B,\nabla^2\rho_B)$ does not depend on $\rho_A$ (to
      obtain the analytic form of $\tilde{T}_s^{nad}[\rho_A,\rho_B]$:
      ($f[\rho_A,\rho_B]\approx \tilde{f}[\rho_B]$)
\end{itemize}
The above criteria are very restrictive and leave us with not many
choices. The last one leads to the following form of the
switching factor:
\begin{eqnarray}
f[\rho_A,\rho_B]\approx \tilde{f}(\rho_B,\nabla\rho_B,\nabla^{2}\rho_B)
\end{eqnarray}
Approximating $f[\rho_A,\rho_B]$ by some function $f(\rho_B)$ 
is one of possible further simplifications. 
It is, however, very unlikely that a $\rho_B$-based switching function
could be universal.
The electron density near the nucleus depends on the effective nuclear
charge $\zeta$~\cite{clementiraimondi1963}
and varies strongly from atom to atom.
It is possible, however, to design an $\zeta$-independent criterion. 
To this end, we consider the 
reduced density gradient ($s_B(\vec{r})$)
 defined as:
\begin{eqnarray}
s_B=\frac{\vert\nabla\rho_B\vert}{2(3\pi^2)^{1/3}\rho_B^{4/3}}
\end{eqnarray}
For $\rho_B$ obtained from hydrogenic orbital $1s$ defined by some
effective nuclear charge $\zeta$, the following $\zeta$-independent
observations can be made:
{\it i}) For $\rho_B^{1s}=2|1s|^2$, 
$s_{r=0}=\left(6\pi\right)^{-1/3}=0.376$, and {\it ii})  
$v_t^{limit}[\rho_B^{1s}]$ changes sign from positive to negative
at $s_{r=2/\zeta}=\exp(4/3)\cdot s_{r=0}=1.426$.
These observations suggest that the switching function can take a very
simple and $\zeta$-independent form 
\begin{eqnarray}
\tilde{f}(\rho_B,\nabla\rho_B)=\tilde{f}(s_B)=\Theta(s_B-s_B^{min})\times\Theta(s_B^{max}-s_B)
\end{eqnarray}
where $\Theta(x)=1$ for $x\ge0$ and $\Theta(x)=0$ for $x< 0$ and
$s_B^{min}=0.376$ and $s_B^{max}=1.426$.

Since Eq.~\ref{ksceda} can be used also to obtain
forces~\cite{Dulak2007}, it is preferable to use a smooth switching  from $0$ to $1$ 
 instead of $\Theta$ in the above definition. 
The simplest
form of such a switching functional has the  Fermi-Dirac statistics
form:
\begin{eqnarray}
\tilde{f}&=&\left({\exp(\lambda(-s_B+s_B^{min}))+1}\right)^{-1}\times
  \left(
1-\left({\exp(\lambda(-s_B+s_B^{max}))+1}\right)^{-1}
\right) \nonumber \\ \label{smoothswitch1}
\end{eqnarray}
where the parameter $\lambda$ determines the smoothness of the switch.

$\lambda=500$ in Eq.~\ref{smoothswitch1} leads to equivalent results to that
obtained with the step function $\Theta(x)$ (differences in dipole
in the range of 10$^{-6}$ Debye and orbital energies in the range of
10$^{-10}$~hartree). All results discussed in this work are obtained
with $\lambda=500$. Smaller values corresponding to even ``softer''
switching can be also used in our numerical implementation but is not 
considered here in order to minimise the use of adjustable parameters.

A switching factor $\tilde{f}$ constructed following the above
restrictions can be used to investigate the importance of enforcing
the exact limit for $v_t[\rho_A,\rho_B](\vec{r})$ at $\rho_A\longrightarrow 0$ and $\int\rho_B
      d\vec{r}=2$ but only in cases where the environment comprises one
      nucleus and two electrons.
Not only the condition  $\int \rho_B d\vec{r}=2$ apply rigorously but
the considerations leading to the $\zeta$-independent values of
$s_B^{max}$ and $s_B^{min}$ apply as well.
Such model systems as Li$^+$-H$_2$O and  Be$^{2+}$-H$_2$O complexes
if the water molecule is considered as subsystem $A$ and the cation as subsystem
$B$ (environment) fall into this category.
At equilibrium geometry 
addition of  $\tilde{f}\cdot v_t^{limit}[\rho_B]$ 
results in a desired effect on calculated properties for these complexes.
The lowest unoccupied embedded orbital associated with subsystem $A$
is indeed localised on the cation and its energy is shifted by 0.262~eV
in the case of Li$^+$ and by 0.830~eV
in the case of Be$^{2+}$.
Addition of $\tilde{f}\cdot v_t^{limit}[\rho_B]$ reduces also the
dipole moment of a water molecule in the vicinity of the cation by
0.071~Debye and 0.409~Debye for  Li$^+$ and  Be$^{2+}$, respectively.
Such noticeable numerical effects obtained for systems, for which the
condition $\int \rho_B d\vec{r}=2$ applies rigorously, indicate clearly
that the exact limit considered might be relevant for practical
calculations where the environment is larger.

\subsection{Fine-tuning of the thresholds in the switching function $\tilde{f}$}

Using only universal parameters $s_B^{min}$ and  $s_B^{max}$ is very
appealing but the reasoning  leading to their numerical values 
does not apply in real systems. 
It allows one to study the importance of the considered exact
conditions but only in particular cases. The approximation to
$v_t[\rho_A,\rho_B](\vec{r})$
defined in Eq.~\ref{NDSD} and using the switching function given in
Eq.~\ref{smoothswitch1} has, therefore, little practical value.
The  construction of the switching function described in the previous
section and choice of the
thresholds $s_B^{min}$ and $s_B^{max}$ in particular can be expected
not to be adequate for other systems for the following reasons:
\begin{itemize}
\item The criterion $0.376\le s_B\le 1.426$ applies at hydrogen nucleus
where addition of $v_t^{limit}[\rho_B]$ is not expected to be needed.
We recall here that it 
is the danger of a collapse of electron density on a doubly occupied
hydrogenic $1s$ orbital provides the physical motivation for 
 introducing the $v_t^{limit}[\rho_B]$.
Moreover, numerical studies on molecular electron densities indicate clearly that
enforcing the local behaviour of the density of the kinetic
energy near nucleus corresponding to the 
von Weizs\"acker expression leads indeed to significant improvements
of the approximation to $T_s[\rho]$ for all nuclei except that of
hydrogen~\cite{Tran2002b}.
\item For heavier atoms, electron density at the nucleus comprises
      contributions from other orbitals than the hydrogenic
      $1s$.
This can lead to the possibility that 
is $v_t^{limit}[\rho_B]$ is negative although  $s_B\le 1.426$.
\item The criterion $0.376\le s_B\le 1.426$ might also be satisfied near stationary
      points of the electron density such as bond midpoints. It is
      very unlikely that the condition $\int \rho_B d\vec{r}=2$ can be
      relevant to any volume element centred on a stationary points.
Therefore addition of  $v_t^{limit}[\rho_B]$ lacks formal
justification there.
Although the condition $0.376\le s_B$
assures that 
 $v_t^{limit}[\rho_B]$ is not added at the stationary point, where
 $\vert\nabla\rho_B\vert=0$, or in the close proximity to it, 
the numerical value of this threshold  requires
 verification in real systems.%
It should be added at this point also that, 
if linear combination of atomic
      orbitals is used to construct embedded orbitals, 
the quality of description of the density
      at the nuclear cusp depends on the used basis set. A weaker
      criterion should be used in practice to assure that the
      $v_t^{limit}[\rho_B]$
is indeed added in the vicinity of the nucleus.
\end{itemize}

We start with the choice made for $s_B^{max}$. 
Numerical analyses in the systems discussed in the next section show that $v_t^{limit}[\rho_B]$ 
is negative locally even if  $s_B$ is smaller than 1.426 for 
heavier nuclei. This suggest that this threshold should be reduced.
The smaller is the value for this threshold the less probably is
inclusion of negative  $v_t^{limit}[\rho_B]$, which is desired from the
point of view of universality of this threshold, but it comes at the
expense of loosing a part of the desired effect at two-electron nuclei
by reducing the range at which addition of  $v_t^{limit}[\rho_B]$ applies.
We use the model system considered in Appendix B
to estimate the effect associated with the reduction of this range.
The underlying assumption leading to the value of 1.426 value is
rigorously true in the model system.  
The desired effect of reducing the charge distribution on top of the
nucleus is achieved mainly by adding  $v_t^{limit}[\rho_B](\vec{r})$ 
very close to the nucleus i.e. where   $s_B<0.6$. 
Increasing further the value of the $s_B$
threshold leads to smaller effect.
Moreover,  adding locally  $v_t^{limit}[\rho_B]$  to the potential
near the nucleus leads to negligible
effect on orbital energies in the system considered in Appendix B
(it reaches a peak of about $10^{-4}$~eV at $s_B=1.426$). 
These results for the model system indicate that any choice for  $0.6<s_B^{max}<1.426$
is acceptable. In  real systems discussed in the next
section,  lowering the threshold from $1.426$ to  to
$0.9$ assures that  $v_t^{limit}[\rho_B]$ is added only if it is positive.

The fine-tuning of
$s_B^{min}$ follows
other considerations. We note that, in the  model system considered in
Appendix B,  $s_B^{min}$
can be reduced even to zero without affecting the results because
 lower values of $s_B$ than 
$0.376$ do not occur near the nucleus.
To make sure that no nuclei is overlooked even if
the chosen atomic basis set in practical calculations is such that the
exact relation $s_B=0.376$ for $r\longrightarrow 0$ cannot be rigorously
satisfied, $s_B^{min}$ is reduced from 0.376 to 0.3. This change leads to
negligible numerical effects if $\rho_B$ corresponds to atomic electron
densities. For molecular $\rho_B$, retaining the criterion based on 
the $s_B^{min}$ is necessary
to avoid
unjustified additions of  $v_t^{limit}[\rho_B]$ near stationary points.

The criteria based only on $s_B^{min}$ and $s_B^{max}$ are not
sufficient if the environment comprises hydrogen atoms because they
are satisfied also at hydrogen nucleus.
To avoid adding the $v_t^{limit}[\rho_B]$ near hydrogens, the
proposed switching function includes additionally the criterion based
on smallness of $\rho_B$. 
It is required that $\rho_B$ is
larger than 
the square of the $1s$
wave function of the hydrogen atom (Z=1) at $r=0$ which equals to
$1/\pi=0.318$.
Concerning $\rho_B^{min}$, increasing the idealised value of
0.318  to even 1 does not affect the results for hydrogen-free systems because 
the density on top of any nucleus, which is heavier than hydrogen, is
at least one order of magnitude larger. Increasing the value
$\rho_B^{min}$
is desired for the same reasons as the ones motivating the decrease of
$s_B^{min}$.
The value of $\rho_B^{min}=0.7$ was arbitrary chosen for practical calculations.

The final form of the ``fine-tuned'' switching function of more
general applicability and used in the subsequent section for studying
the importance of imposing the considered exact limit in real systems
take the following form:
\begin{eqnarray}
\tilde{f}&=&\left({\exp(\lambda(-s_B+s_B^{min}))+1}\right)^{-1}\times
  \left(
1-\left({\exp(\lambda(-s_B+s_B^{max}))+1}\right)^{-1}
\right) 
\nonumber\\
&\times&  \left({\exp(\lambda(-\rho_B+\rho_B^{min}))+1}\right)^{-1}\label{smoothswitch}
\end{eqnarray}
where $s_B^{min}=0.3$,  $s_B^{max}=0.9$, $\rho_B^{min}=0.7$.

Eqs.~\ref{vtsnad_gea0}, \ref{vsnadw}, \ref{NDSD},
and \ref{smoothswitch} define 
the potential which will be referred to as
$\tilde{v}_t^{NDSD}[\rho_A,\rho_B]$, 
(Non-Decomposable approximation using first- and  Second
Derivatives of $\rho$).
The notion of non-decomposability is brought up here because the the second term in Eq.~\ref{NDSD}
does not have the form of a difference between two functional
derivatives of some common explicit density functional
$\tilde{T}_s[\rho]$.
The analytic expression for $\tilde{T}_s^{nad(NDSD)}[\rho_A,\rho_B]$, which yields
$\tilde{v}_t^{NDSD}[\rho_A,\rho_B]$
after functional differentiation with respect to $\rho_A$
can be easily constructed.
Its decomposable component is given in Eq.~\ref{tsnad_gea0}
and 
the non-decomposable  $v_t^{limit}[\rho_B]$ component is $\rho_A$
independent. The functional generating  $v_t^{NDSD}[\rho_A,\rho_B]$
reads therefore:
\begin{eqnarray}
\tilde{T}_s^{nad(NDSD)}[\rho_A,\rho_B]&=&
C_{TF}\int \left(
\left(\rho_A+\rho_B\right)^{5/3}
-\rho_A^{5/3}
-\rho_B^{5/3}
\right)d\vec{r} \\
&+&
\int
f(\rho_B,\nabla\rho_B)\cdot \rho_A(\vec{r}) \tilde{v}_t^{limit}[\rho_B](\vec{r}) d\vec{r}
+ C[\rho_B] \nonumber
\end{eqnarray}
where $C[\rho_B]$ is  $\rho_A$-independent. To assure the
proper dissociation limit $C[\rho_B]$ must vanish.

\section{Numerical validations}

\subsection*{Procedure to  analyse $\tilde{v}_t$-generated
  errors} 

In practical applications of Eq.~\ref{ksceda}, the  results depend
on $\rho_B$ as well as on the used approximation to
$v_t[\rho_A,\rho_B](\vec{r})$.
As far as  the quality of the used approximation to
 $v_t[\rho_A,\rho_B](\vec{r})$ is concerned, 
a general procedure was proposed in
 one of our earlier works~\cite{bernard2008,Wesolowski1996}.
Its principal element is the comparison between numerical values 
of the calculated
property (energy components, dipole moments, total electron density,
etc.) obtained from two fully variational formal frameworks: that of
Cortona~\cite{Cortona1991} and that of Kohn and Sham~\cite{KS1965}.
Results obtained from both frameworks are not exact but the difference between them can be 
attributed only to the approximation used for
$v_t[\rho_A,\rho_B](\vec{r})$ if  all
technical parameters  (approximation
to the exchange-correlation functional, basis sets for expanding
orbitals,  algorithms to
calculate used matrix elements) are the same.
We point out here, however, that direct comparisons between the total electron densities
derived from Kohn-Sham- and Cortona's calculations%
are cumbersome because these quantities are local.
In practice, it is more convenient to use global quantities (norm of
the difference between these densities, or selected observables)
in such analyses (see the next section). 

To obtain the pair of electron densities $\rho_A$ and $\rho_B$
which minimises the total energy in Cortona's type of calculations, 
a self-consistent super cycle
  of embedding calculations ({\it freeze-and-thaw} cycle) 
is performed.
At each iteration, Eq.~\ref{ksceda}
are solved. In the subsequent iteration, $\rho_A$ and $\rho_B$
exchange their role in Eqs.~\ref{ksceda}.%
The {\it freeze-and-thaw} iterations continue until self-consistency. 
In the end, a pair of electron densities ($\rho_A^0$ and
$\rho_B^0$)
and the corresponding two sets of embedded orbitals is obtained.
Obviously,  the notion of {\it embedded system} and its {\it environment}
becomes meaningless because both subsystems are treated on the equal
footing.
{\it Freeze-and-thaw} calculations are conducted in practice for small
model systems to 
validate the used $\rho_B$ in large scale multi-level numerical
simulations~\cite{Neugebauer_excitations2} or, as it is made in the present work, to 
asses the used approximation to $v_t[\rho_A,\rho_B](\vec{r})$. 
\subsection*{The  $\tilde{v}_t$-generated errors in the complexation induced dipole moments due to
  violation of the limit for $v_t$ at $\rho_A\longrightarrow 0$ and $\int\rho_B
      d\vec{r}=2$.}
 $\tilde{v}_t$-generated errors in complexation induced
dipole moments can be expected to be strongly affected by the local
behaviour of the used  $\tilde{v}_t[\rho_A,\rho_B](\vec{r})$ near
nuclei.  Lack of sufficient repulsion near the nucleus
might lead to an artificial transfer of electron density between
subsystems reflected in the numerical values of the dipole moment.
Therefore, this quantity was  chosen for the analysis of the 
 $\tilde{v}_t$-generated errors in a representative set of intermolecular complexes
including charged, polar and non-polar ones at their equilibrium geometries.
For  key details of the numerical implementation of the relevant
equations, see  Ref.~\cite{computationaldetails}.

Tables~\ref{dipolemoments.tab} and \ref{dipolemoments_charged.tab}
collect the complexation induced dipole moments in neutral or charged
complexes, respectively.
First of all, switching on the $v_s^{limit}[\rho_B](\vec{r})$ term
decreases the $\tilde{v}_t$-generated errors in each of
the considered cases.
For systems comprising neutral subsystems the effect on the errors are
negligible. This indicated that the origin of the errors lies not in
the violation the condition considered in this work.
For systems comprising charged components, the effect of imposing the
considered limit is evident. The errors are invariably reduced. The
reduction of the relative errors depends on the system 
from such a case as Li$^+$-H$_2$O  (from 9.8\% to 7.9\%) to a
reduction by factor 2 or 3 in  Na$^+$-Br$^-$ and (from 1.4\% to 0.7\%)
and Na$^+$-H$_2$O (from 0.23\% to 0.07\%).
Similarly as in the group of complexes formed by neutral molecules,
the origin for the remaining errors lies somewhere else.
The above numerical examples lead to the following principal
conclusions:
\begin{itemize}
\item Violation of the limit for $v_t[\rho_A,\rho_B](\vec{r})$ at $\rho_A\longrightarrow 0$ and $\int\rho_B
      d\vec{r}=2$ contributes to the overall error in the calculated
      quantities but this contribution varies from one system to
      another. It is rather negligible for complexes formed by neutral
      components. It is numerically significant for complexes
      comprising charged components.
\item Our simple strategy to impose the considered limit locally in
      the vicinity of nuclear cusps leads invariably to reduction of
      errors. Therefore it can be used generally as correction to any
      approximation violating the above limit.
\item The construction of the approximation obeying the considered
      limit, and the used switching criteria in particular, corresponds
      to a real case where a distant nucleus is surrounded
      by a frozen-density shell comprising two electrons (He, Li$^+$,
      Be$^{2+}$, etc.).
      The fact that the errors are reduced also for systems, where
      these idealised conditions do not apply,
      indicates that the considered condition is important and should
      be taken into account in construction of approximations to 
      ${v}_t[\rho_A,\rho_B](\vec{r})$.
\item For systems, where  $\tilde{v}_t$-generated errors were not reduced
      by imposing the considered exact limit,  their
      origin must be looked for somewhere else.
\end{itemize}

\subsection*{The effect of imposing the limit for $v_t$ at $\rho_A\longrightarrow 0$ and $\int\rho_B
      d\vec{r}=2$  on orbital energies}

In this section, we analyse the complexation induced shifts of orbital energies derived
using  the approximated potential
$\tilde{v}_t^{NDSD}[\rho_A,\rho_B](\vec{r})$ considered in the previous section.
Opposite to the dipole moments discussed previously, direct
comparisons between the calculated shifts and the corresponding reference data are
less straightforward.  
However, we investigate the numerical effect associated with imposing the exact
limit for $v_t[\rho_A,\rho_B](\vec{r})$ at $\rho_A\longrightarrow 0$ and $\int\rho_B
      d\vec{r}=2$  in view of the numerical practice which indicates that
shifting the levels of unoccupied orbitals localised in the
environment would be desirable.
Conventional decomposable approximations to
${v}_t[\rho_A,\rho_B](\vec{r})$ lead to artificially low levels of
unoccupied orbitals in the environment which
 might cause unphysical effects such as charge-transfer between
subsystems~\cite{dulak_private} or erroneous other
observables~\cite{jacob}.

The numerical results for a model system considered in Appendix B show
that
inclusion of the $v_t^{limit}[\rho_B]$ term everywhere where
$0.6\le s_B \le 0.9$ into the effective
embedding potential leads to a positive shift of the energy level of
the unoccupied orbital.
In real intermolecular systems, the conditions considered in Appendix B
($\rho_A\longrightarrow 0$ and $\rho_B=2|1s|^2$) are not satisfied.
The subsystems are in finite separation and the use of
atom-centred basis sets for each subsystem, which include all atoms
(supermolecular expansion labelled as KSCED(s) in
Ref.~\cite{Wesolowski1997}), results in the fact that 
$\rho_A$ can be significant at a nucleus associated with subsystem
$B$. For the same reasons and the fact that the considered nuclei
include also atoms with occupied $2s$ shell, also the second assumption is
 not satisfied rigorously.
Therefore, it is useful to verify in practice to which extend inclusion of the
$v_t^{limit}[\rho_B]$ term
affects the orbital levels if these asymptotic conditions 
do not apply.

Tables~\ref{lueo_shifts_1} and \ref{lueo_shifts_3} collect the values
of energy
levels corresponding to the lowest lying  orbital localised mainly 
in the environment for the previously considered complexes.
The calculations are made not in the end of the {\it freeze-and-thaw} cycle
but for the same $\rho_B$ obtained from Kohn-Sham calculations for the
isolated subsystem B.
Including
$v_t^{limit}[\rho_B]$
leads to  the shifts of the
energy levels of unoccupied orbitals of the magnitude which significantly larger 
than that in the model system.
It is a very desired effect of the new approximation.

In the  Li$^+$-H$_2$O case discussed in
Ref.~\cite{jacob},
the energy of the lowest unoccupied embedded orbital localised on
Li$^+$ crosses that of the highest occupied embedded orbital localised on
H$_2$O at the intermolecular distance of 13~\AA~if Eq.~\ref{vtsnad_gea0}
is used for $v_t[\rho_A,\rho_B](\vec{r})$. At larger separations,  the self-consistent procedure  
to solve  Eqs.~\ref{ksceda} does not converge due to localisation of
the highest occupied embedded orbital which jumps between subsystems
in subsequent iterations. Addition of the $v_t^{limit}[\rho_B]$ term
shifts the energy of the unoccupied embedded orbital localised at
Li$^+$. As the result, no crossing of levels occurs 
even at  intermolecular separations as large as 18~\AA.
The occupied levels are affected less strongly (see
Tables~\ref{hoeo_shifts_1} and \ref{hoeo_shifts_3}).

\section{Discussion}

In principle, any decomposable
approximation can be used as the first term of Eq.~\ref{NDSD} instead
of the term derived from Thomas-Fermi functional.
In the approximation introduced in this work, the decomposable
component of $\tilde{v}_t^{NDSD}[\rho_A,\rho_B]$ is gradient-free and does not
contribute to the asymptotic  local behaviour of $v_t[\rho_A,\rho_B](\vec{r})$ at the nuclei.
Only the second term enforces the desired  behaviour.
Should a gradient-dependent alternative for the first term be considered,
a proper care should be taken to avoid double-counting of $v_t^{limit}[\rho_B]$
 at a nucleus. For instance, $\tilde{v}_t^{GEA2}[\rho_A,\rho_B]$ comprises
already 1/9 of $v_t^{limit}[\rho_B]$. 
To verify  whether further improvements are possible following this lines,
two approximations were considered by replacing $\tilde{v}_t^{TF}[\rho_A,\rho_B]$ in
Eq.~\ref{NDSD} by either  $\tilde{v}_t^{GEA2}[\rho_A,\rho_B]$~\cite{WW1993} or
$\tilde{v}_t^{GGA97}[\rho_A,\rho_B]$~\cite{Wesolowski1997a}.
Including the
$v_t^{limit}[\rho_B]$ term into $\tilde{v}_s[\rho_A,\rho_B](\vec{r})$
brings improvements in both cases (see
Table~\ref{other.decomposable}).
This indicates $v_t^{limit}[\rho_B]$ should be enforced universally on
any approximation to  ${v}_s[\rho_A,\rho_B](\vec{r})$.
Compared to the results obtained using  $\tilde{v}_s^{NDSD}[\rho_A,\rho_B](\vec{r})$
discussed earlier in this work, the two alternative non-decomposable
approximations ${v}_s[\rho_A,\rho_B](\vec{r})$ are not better.
We recall here the main reasons for singling out
$\tilde{v}_t^{GGA97}[\rho_A,\rho_B]$)
among other decomposable ones which depend explicitly on
densities $\rho_A$ and $\rho_B$ as well as 
their first- and second derivatives.
\begin{itemize}
\item  $\tilde{v}_t^{GEA2}[\rho_A,\rho_B]$ obtained from the second-order
      gradient-expansion approximation 
      leads typically to worse results than
      that obtained from zeroth
      order~\cite{bernard2008,Wesolowski1997} indicating that the
contribution to $v_t[\rho_A,\rho_B]$ due to the second term in Eq.~\ref{tsnad_gea2}
is erroneous.
The deterioration of the results is pronounced the most at very small
overlaps between $\rho_A$ and $\rho_B$. Note that in the present work
this flaw of  $\tilde{v}_t^{GEA2}[\rho_A,\rho_B]$ manifests itself 
in the absence of convergent solutions of Eq.~\ref{ksceda} in
Na$^+$-Cl$^-$ and Li$^+$-H$_2$O cases (see Table~\ref{other.decomposable}).
\item   $\tilde{v}_t^{GGA97}[\rho_A,\rho_B]$ was introduced as a
      pragmatic solution replacing  $\tilde{v}_t^{GEA2}[\rho_A,\rho_B]$.
Due to its analytic form, the gradient-dependent contribution 
disappears at small overlaps between $\rho_A$ and $\rho_B$.
\item The functional
$\tilde{T}^{LC94}[\rho]$ generating the decomposable approximated
potential $\tilde{v}_s^{GGA97}[\rho_A,\rho_B]$
is known to be a very good approximation to $T_s[\rho]$.
\end{itemize}
The above reasons for singling out
 $\tilde{v}_t^{GGA97}[\rho_A,\rho_B]$ are not applicable for the non-decomposable
construction presented in this work.
In $\tilde{v}_t^{NDSD}[\rho_A,\rho_B]$ the problematic second-order
term
lying at the origin of flaws of  $\tilde{v}_t^{GEA2}[\rho_A,\rho_B]$
is either present where it is needed to assure the correct asymptotic
limit (i.e. in the vicinity of nuclei)
or absent.
Numerical results collected in 
Tables~\ref{dipolemoments.tab}, \ref{dipolemoments_charged.tab}, and
\ref{other.decomposable} 
support fully the above formal reasons to consider
$\tilde{v}_t^{NDSD}[\rho_A,\rho_B]$
as the successor of  $\tilde{v}_t^{GGA97}[\rho_A,\rho_B]$.
\section{Conclusions}

Enforcing the considered exact limit for
${v}_t[\rho_A,\rho_B](\vec{r})$,
as it is made in the proposed approximation, leads to a
significant reduction of  $\tilde{v}_t$-generated errors for charged
systems.
Errors in the complexation indiced dipole moments are reduced 
by more than 50\% in some cases. For neutral systems, reduction of error takes also place but its
magnitude is typically negligible.
This indicates that one of important sources of inaccuracies in the
conventional (i.e. decomposable and gradient-expansion based)
 approximations to ${v}_t[\rho_A,\rho_B](\vec{r})$ was
identified.
The origin of the remaining contributions  $\tilde{v}_t$-generated errors
lies probably somewhere else.
The analytic form of the component of the local embedding potential
enforcing the correct considered limit is, indeed, an approximation to
${v}_t[\rho_A,\rho_B](\vec{r})$ because 
 its position dependency is indirect - through the density and its gradient.
The function which was used to switch on the exact limit 
was designed based on the analysis of model
system of relevance for elements of the first-, second- and the third
period.
Using the  local potential introduced in this work 
($\tilde{v}_t^{NDSD}[\rho_A,\rho_B](\vec{r})$ given in Eq.~\ref{NDSD}) as an alternative to
potentials derived using conventional strategy of deriving it from analytic form 
of functionals based on low-order terms 
in the gradient expansion of $T_s[\rho]$~\cite{Kirzhnitz} is
 recommended for the following reasons:\\
{\it i}) {\bf Formal:} We believe that a proper strategy to improve
approximations  to functionals in density functional theory should proceed by imposing
the most relevant exact conditions and $\tilde{v}_t^{NDSD}[\rho_A,\rho_B](\vec{r})$
was constructed in this way.\\
{\it ii}) {\bf Practical:} The numerical results reported in  this work show
indeed that imposing this condition improves the obtained electron
density and that the improvement varies from negligible to significant
depending on the system.\\
{\it iii}) {\bf Numerical:} Evaluating
$\tilde{v}_t^{NDSD}[\rho_A,\rho_B](\vec{r})$
involves the same quantities as evaluating its counterparts derived
from gradient-expansion- (up to second order) and so called generalised
gradient approximations to $T_s[\rho]$.

Concerning the area of applicability of
$\tilde{v}_t^{NDSD}[\rho_A,\rho_B](\vec{r})$,
it should be underlined that 
some arbitrary choices were made concerning the criteria for
``detecting'' the vicinity of a nucleus based only on electron density
$\rho_B$ and its derivatives 
in the construction of this approximation.%
The chosen criteria are most adequate for such nuclei, where  the
total electron density is dominated by the hydrogenic $1s$ orbital.

The local potential $\tilde{v}_t^{NDSD}[\rho_A,\rho_B](\vec{r})$ 
introduced in this work is
intrinsically  non-decomposable.
Although the analytic form of the functional
$\tilde{T}_s^{nad(NDSD)}[\rho_A,\rho_B]$  and the potential
$\tilde{v}_t^{NDSD}[\rho_A,\rho_B](\vec{r})$ 
are given in this work, 
the analytic form of 
neither $\tilde{T}_s^{NDSD}[\rho]$ nor
$\frac{\delta \tilde{T}_s^{NDSD}[\rho]}{\delta\rho}$ is available. 
Therefore, the introduced here non-decomposable strategy  can be seen as the first attempt
to decouple the search for $\tilde{T}_s[\rho]$ and its functional
derivative, which are needed
in orbital-free calculations, from the search for an adequate  approximation
for the kinetic-kinetic-energy dependent component of the effective
orbital-free embedding potential.
Opposite to orbital-free strategy~\cite{ThomasFermi},  neither $\tilde{T}_s[\rho]$
nor its functional derivative are needed in methods applying
orbital-free effective embedding potential given in
Eq.~\ref{kscedembpot}.

Finally,  imposing the exact limit for $v_t[\rho_A,\rho_B](\vec{r})$ at $\rho_A\longrightarrow 0$ and $\int\rho_B
      d\vec{r}=2$  leads to  shifts of the level of the unoccupied
orbitals localised in the environment. Such shift is strongly desired 
in view of the earlier reports on possible practical 
inconveniences resulted from artificially low position of such levels 
when approximations not taking into account the cusp condition are
used~\cite{jacob,dulak_private}.

{\it Acknowledgement:} This work was supported by Swiss National Science Foundation.

\newpage
\section*{Appendix A}
\subsubsection*{Orbital-free embedding potential for
  $\rho_A\longrightarrow 0$ and $\rho_B=2\vert 1s\vert^2$}
$\;$ \\
For  small $\delta\rho$ such that $\delta\rho\longrightarrow 0$,\\
\begin{eqnarray}
\delta T_s^{nad}[\rho_A,\rho_B]&=&T_s^{nad}[\rho_A+\delta \rho,\rho_B]-T_s^{nad}[\rho_A,\rho_B]
 \\
 \nonumber \\
&=&T_s[\rho_A+\delta\rho+\rho_B]-T_s[\rho_A+\rho_B]-T_s[\rho_A+\delta\rho] + T_s[\rho_A] \nonumber \\
 \nonumber \\
&=& \int \left.\frac{\delta T_s[\rho]}{\delta
    \rho}(\vec{r})\right|_{\rho=\rho_A+\rho_B}
\delta{\rho}(\vec{r})d\vec{r} 
- \int \left.\frac{\delta T_s[\rho]}{\delta \rho}(\vec{r})\right|_{\rho=\rho_A} \delta{\rho}(\vec{r})d\vec{r} 
 + O(\delta^2\rho) \nonumber  
\end{eqnarray}
If also $\rho_A$ is small i.e. $\rho_A\longrightarrow 0$,\\
\begin{eqnarray}
T_s^{nad}[\rho_A+\delta \rho,\rho_B]-T_s^{nad}[\rho_A,\rho_B] =%
\int \left.\frac{\delta T_s[\rho]}{\delta \rho}(\vec{r})\right|_{\rho=\rho_B} \delta{\rho}(\vec{r})d\vec{r} + O(\delta^2\rho) \nonumber \\ 
\end{eqnarray}
Therefore,
\begin{eqnarray}
\left.\frac{\delta T_{s}^{nad}[\rho,\rho_{B}]}{\delta \rho}(\vec{r})\right|_{\rho\longrightarrow 0} \approx
\left.\frac{\delta T_{s}[\rho]}{\delta
    \rho}(\vec{r})\right|_{\rho\longrightarrow \rho_B} \label{approx1}\\ \nonumber
\end{eqnarray}
The above result that the kinetic-energy component of $v_{emb}^{KSCED}$ is
just the functional derivative of the $T_s[\rho]$ calculated for
$\rho=\rho_B$ makes it possible to express it analytically for any
$\rho_B$ which comprises just two electrons.
For one-electron and two-electron- spin-compensated systems the exact
expression reads~\cite{Weiz}:
\begin{eqnarray}
T_{s}[\rho]=T^W_{s}[\rho]=\int \frac{1}{8}\frac{\nabla \vert\rho\vert^2}{\rho}d\vec{r}\;\;\mathrm{for}\;\;\int\rho d\vec{r}=2
\end{eqnarray}
Therefore,
\begin{eqnarray}
\left.\frac{\delta T^W_{s}[\rho]}{\delta
    \rho}\right|_{\rho=\rho_B}=\frac{1}{8}\frac{\vert
  \nabla \rho_B\vert^2}{\rho_B^2} -
\frac{1}{4}\frac{\nabla^2\rho_B}{\rho_B}\;\;\mathrm{if}\;\;\int\rho_B d\vec{r}=2 \label{derivativeW} \\ \nonumber
\end{eqnarray}
Using Eq.~\ref{derivativeW} in Eq.~\ref{approx1} leads to
the asymptotic form of the kinetic energy component of $v_{emb}^{KSCED}$
in the case where $\rho_A$ and $\rho_B$ do not overlap significantly
and $\rho_B$ is a two-electron spin-less electron density reads:
\begin{eqnarray}
\left.\frac{\delta T_{s}^{nad}[\rho,\rho_{B}]}{\delta
    \rho}\right|_{\rho=\rho_A\longrightarrow 0,\int\rho_B d\vec{r}=2}
= \frac{1}{8}\frac{\vert \nabla \rho_B\vert^2}{\rho_B^2} - \frac{1}{4}\frac{\nabla^2\rho_B}{\rho_B}
 \label{assymptotic_vtnad}
\end{eqnarray}
$\rho_B$ representing a doubly occupied
hydrogenic $1s$ function ($1s=\sqrt{\zeta^3/\pi}\cdot\exp(-\zeta r)$) reads:
\begin{eqnarray}
\rho_B^{1s}(\vec{r})=\rho_B^{1s}(r)=2\cdot \zeta^3/\pi\cdot \exp\left(-2\zeta r\right), \label{defrhob}
\end{eqnarray}
For $\rho_B^{1s}$, Eq.~\ref{assymptotic_vtnad} leads to the following potential:
\begin{eqnarray}
\left.\frac{\delta T_{s}^{nad}[\rho,\rho_{B}]}{\delta
    \rho}\left(\vec{r}\right)\right|_{\rho=\rho_A\longrightarrow 0,\rho_B = \rho_B^{1s}}=\frac{\zeta}{r} -
\frac{\zeta^2}{2}\label{exactpotential} \\ \nonumber
\end{eqnarray}
The potential given in Eq.~\ref{exactpotential} is repulsive for $r <
\frac{2}{\zeta}$. For hydrogenic densities $\rho_B^{1s}$, the reduced
density gradient equals to a $\zeta$-independent value of 1.426.
Near the nuclear cusp, therefore,  a pair of
electrons on the $1s$ shell provides a local repulsive potential which
compensates the Coulomb attraction due to the nuclear
charge. Note that the effective nuclear charge $\zeta$
  for the most tightly bound orbital
 ($1s$) in a multi-electron atom  is smaller than
the charge of the corresponding nucleus ($Z$)~\cite{clementiraimondi1963}. 
As a consequence, the
compensation is perfect only for one-electron hydrogenic systems. 

\newpage
\section*{Appendix B}
\subsubsection*{The effect of approximations to
  $v_t[\rho_A,\rho_B](r)$ 
on the energies of bound states localised in the environment far from
subsystem $A$}

We consider numerical solutions of the Schr\"odinger
equation for one electron in the spherically symmetric potential
which takes the general from given in
Eqs.~\ref{kscedpot1}-\ref{kscedembpot}.
The analysed potential corresponds to
$v_{eff}^{KS}[\rho_A;\vec{r}]=0$, $v_{ext}^B(\vec{r})=-\frac{Z}{r}$,
and  $v_{emb}^{KSCED}[\rho_A,\rho_B;\vec{r}]$ defined in
Eq.~\ref{kscedembpot} for $\rho_A\longrightarrow 0$ and
$\rho_B=\rho_B^{1s}$. Such a case
represents a local potential around a nucleus with the charge $Z$ localised in the
environment far from from the investigated subsystem $A$ and $\rho_B$
comprising entire contribution from doubly occupied $1s$ shell
centred on this nucleus. 
Dirac's exchange energy expression~\cite{Dirac} is used to derive the
exchange-correlation component of $v_{emb}^{KSCED}[\rho_A,\rho_B;\vec{r}]$.
The solutions of one-electron Schr\"odinger equation for such
potential is  obtained by radial quadrature  described in
Refs.~\cite{Fornberg1998,Weber2004} implemented numerically using
MATLAB2006 environment~\cite{MATLAB2006}.
Finite difference approximation and the
matrix representation with 127 radial points is used. 
The above model system is used to investigate the effect of approximations to  $v_t[\rho_A,\rho_B](r)$ 
on the lowest energy level.

Using such approximate potential $\tilde{v}_t[\rho_A,\rho_B](r)$,
which is finite at nuclear cusp, might lead to appearance of
artificially stabilised bound states due to improper balance between
the nuclear attraction,  classical electron-electron repulsion, which
are both descried exactly and improperly behaving
$\tilde{v}_t[\rho_A,\rho_B](r)$.
This imbalance does not cause qualitative problems in the  Z=2
(helium atom) case. No bound states occur
if $v_t[\rho_A,\rho_B](r)$ is approximated  by: 
\begin{eqnarray}
\tilde{v}_t^{model0}&=&\left.\frac{\delta
    T_s^{nad(TF)}[\rho,\rho_B]}{\delta\rho}\right|_{\rho=\rho_A\longrightarrow 0} \label{vt_model1}\\
&=&
\frac{5}{3}C_{TF}
\rho_B^{2/3}\nonumber
\end{eqnarray}
derived from the uniform-electron gas expression for $v_t$ given in
Eq.~\ref{vtsnad_gea0}.

For $Z=3$ (i.e. Li$^+$), however, $\tilde{v}_t^{model0}$
leads to a bound state of the energy -0.2096654 hartree. 
This value will be used as a reference for analysis of other
approximations to $v_t[\rho_A,\rho_B](\vec{r})$.\\

The expression derived from the second-order gradient approximation to $\frac{\delta
  T_s^{nad}[\rho_A,\rho_B]}{\delta\rho_A}$ reads:\\
\begin{eqnarray}
\tilde{v}_t^{model1}&=&\left.\frac
{\delta T_{s}^{nad(GEA2)}[\rho,\rho_{B}]}
{\delta \rho}\right|_{\rho=\rho_A\longrightarrow 0}\label{vtnad_gea2} \\
&=&
\frac{5}{3} 
C_{TF}
\rho_B^{2/3}
-\frac{1}{72} 
\frac{\nabla^2\rho_B}{\rho_B}
+\frac{1}{144}
\frac
{\vert \nabla \rho_B \vert^2}
{\rho_B^2} \nonumber
 \\ \nonumber
\end{eqnarray}
leads to a further lowering of the energy to -0.611935 hartree.

\begin{figure}[h]             %
\centerline{\psfig{file=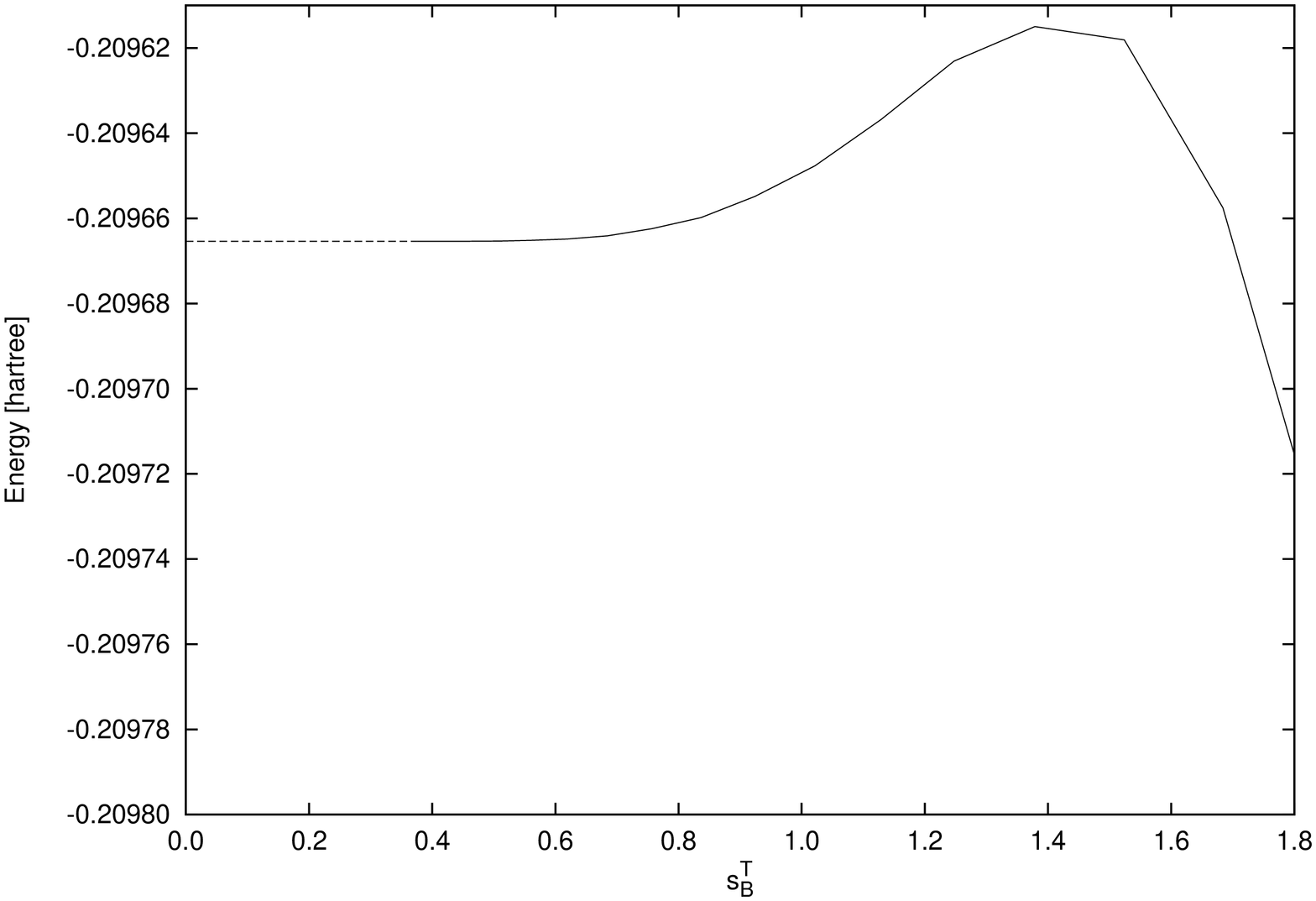,width=3.3in,angle=-90}}
\caption{
The ground-state energy level in the model system 
for different values of the threshold $s_B^T$ corresponding to the  range
 $r^{T}$
of the $\frac{1}{8}\frac{\vert \nabla \rho_B\vert^2}{\rho_B} -
\frac{1}{4}\frac{\nabla^2\rho_B}{\rho_B^2}$ term 
 in Eq.~\ref{vtnad_model2}.
For $s_B^T<0.376$,  the corresponding $r^{T}$ does not exists in the
model system and the results obtained without this additional term are
shown. 
\label{levels_sB}}                                                                                  
\end{figure}

\begin{figure}[h]             %
\centerline{\psfig{file=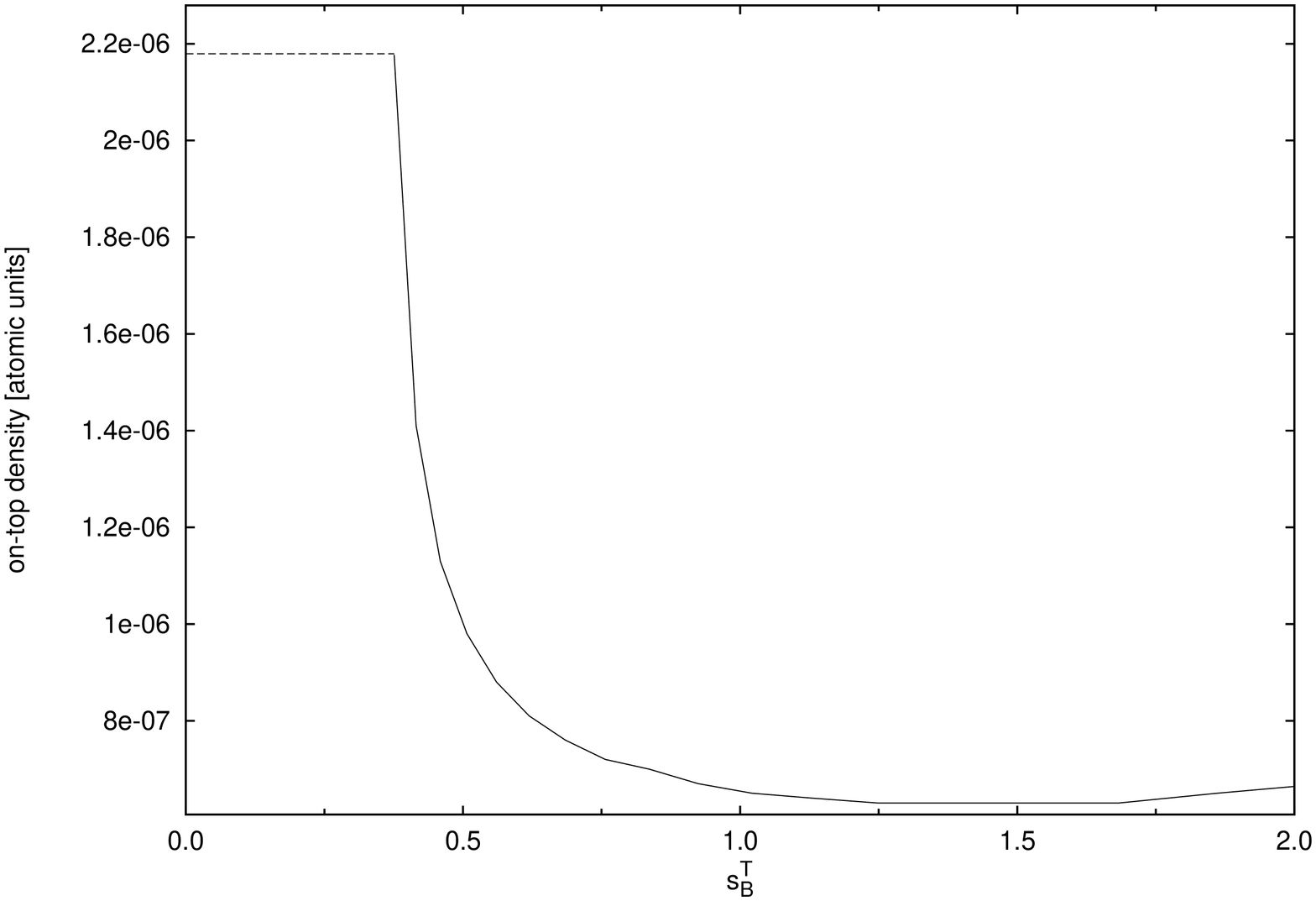,width=3.3in,angle=-90}}
\caption{
The electron density at $r=0$  in the model system 
for different values of the threshold $s_B^T$ corresponding to the  range
 $r^{T}$
of the $\frac{1}{8}\frac{\vert \nabla \rho_B\vert^2}{\rho_B} -
\frac{1}{4}\frac{\nabla^2\rho_B}{\rho_B^2}$ term 
 in Eq.~\ref{vtnad_model2}.
For $s_B^T<0.376$,  the corresponding $r^{T}$ does not exists in the
model system and the results obtained without this additional term are
shown. 
\label{ontopdensity_sB}}                                                                                  
\end{figure}

In the following part, other potentials will be considered, for which
 a local non-decomposable contribution
is added to that given in Eq.~\ref{vt_model1} near the nucleus.
\begin{eqnarray}
\tilde{v}_t^{model2}(r)&=& \frac{5}{3}C_{TF}
\rho_B^{2/3} \label{vtnad_model2}\\
&+&  \left\lbrace 
\begin{array}{l}
 \frac{1}{8}\frac{\vert \nabla \rho_B\vert^2}{\rho_B} -
\frac{1}{4}\frac{\nabla^2\rho_B}{\rho_B^2}=
\frac{\zeta}{r}-
\frac{\zeta^2}{2}\;\;\mathrm{for}\;r\le r^{T}\\
 \\
0 \;\;\mathrm{for}\;r > r^{T}\\
\end{array}
\right. \nonumber
\end{eqnarray}
Figure~\ref{levels_sB} 
shows  that, compared to the decomposable result, 
 the addition of a non-decomposable component 
destabilises 
 the energy but only for such small values
of $r$ at which $s_B \le 1.6$.
The maximal destabilisation occurs at 
$s_B=1.426$ in line with the change of sign of the added term.
A similar picture emerges from the analysis of 
the  electron density at $r=0$.
Comparisons of Figures~\ref{levels_sB} and
\ref{ontopdensity_sB} reveals that
addition of the non-decomposable term to the effective potential 
affects the orbital energies and electron density in a different
manner. Even very close to the nucleus this addition reduced electron
density without affecting the orbital energy noticeably.
From the point of view of choosing $s_B^{max}$ determining the range of
this additional potential it is worthwhile to notice that the limit
$s=1.426$ should not be exceeded. The energy level starts a rapid
descend
and the on-top density starts to rise again at larger values of
$s_B^T$.
As far as the lower limit for $s_B$ is concerned, it should not be
smaller than about 0.5 because points at which  $s_B < 0.5$ influence
significantly the charge density at the nucleus.

\newpage

\begin{table}[h]
\centering
\caption{
Total dipole moment ($\mu$ in Debye) obtained from the {\it freeze-and-thaw}
calculations using 
two approximations $v_t[\rho_A,\rho_B](\vec{r})$.
The target Kohn-Sham results are given for comparison. 
Only dipole
moment components along the principal axis connecting the subsystems are
given.
Relative percentage errors of the dipole moments derived from
subsystem based calculations ($\frac{\Delta \mu^{subsystem}-\Delta
  \mu^{Kohn-Sham}}{\Delta \mu^{Kohn-Sham}}100\%$)
are given in parentheses.
}
$\;$\\
\begin{tabular}{cc|rr|r}
\hline
\hline
  &  &\multicolumn{2}{c|}{freeze-and-thaw}&Kohn-Sham\\
\hline
&&& &\\
  &  &Eq.~\ref{vtsnad_gea0}&Eq.~\ref{NDSD}&\\
\hline
$A$ & $B$&\multicolumn{3}{c}{$\mu$}\\
\hline
 Li$^+$  & F$^-$  & 5.429  &5.542 & 6.020\\
   &   & (9.8)  & (7.9)&\\
 Li$^+$  & Cl$^-$  & 5.768  &5.938 &6.828\\
   &   & (15.5)  & (13.0)& \\
 Li$^+$  & Br$^-$  & 5.706  &5.899&7.033\\
   &   & (18.9)  & (16.1)&\\
 Na$^+$  & F$^-$  & 7.576 & 7.612& 7.697\\
   &   & (1.6)  & (1.1)& \\
 Na$^+$  & Cl$^-$  & 8.509 & 8.563& 8.608\\
   &   & (1.2)  & (0.5)& \\
 Na$^+$  & Br$^-$  & 8.637  &8.698 &8.758\\
   &   & (1.4)  & (0.7)& \\
 Be$^{2+}$  & O$^{2-}$  & 4.225  &4.363 & 6.120\\
   &   & (31.0)  & (28.7)& \\
 Mg$^{2+}$  & O$^{2-}$  & 6.741  &6.822 & 6.988\\
   &   & (3.5)  & (2.4)& \\
 H$_2$O  & H$_2$O  & 2.669  & 2.670&2.779\\
   &   & (3.9)  & (3.9)&\\
 HF  & HF  & 3.000  &3.000 & 3.090\\
   &   & (2.8)  & (2.8)& \\
  He  & CO$_2$  & 0.014  &0.014 & 0.013\\
   &   & (-8.6)  & (-8.4)&\\
  Ne  & CO$_2$  & 0.027  &0.027 & 0.025\\
   &   & (-7.6)  & (-7.6)&\\
\hline
\hline
\end{tabular}
\label{dipolemoments.tab}
 \\
\end{table}

\newpage

\begin{table}[h]
\centering
\caption{
Total dipole moment ($\mu$ in Debye) obtained from the {\it freeze-and-thaw}
calculations using 
two approximations $v_t[\rho_A,\rho_B](\vec{r})$.
The target Kohn-Sham results are given for comparison. 
Only dipole
moment components along the principal axis connecting the subsystems are
given.
Relative percentage errors of the dipole moments derived from
subsystem based calculations ($\frac{\Delta \mu^{subsystem}-\Delta
  \mu^{Kohn-Sham}}{\Delta \mu^{Kohn-Sham}}100\%$)
are given in parentheses.
}
$\;$\\
\begin{tabular}{cc|rr|r}
\hline
\hline
  &  &\multicolumn{2}{c|}{freeze-and-thaw}&Kohn-Sham\\
\hline
&&& &\\
  &  &Eq.~\ref{vtsnad_gea0}&Eq.~\ref{NDSD}&\\
\hline
$A$ & $B$&\multicolumn{3}{c}{$\mu$}\\
\hline
 Li$^+$  & H$_2$O  & 5.298 & 5.353& 5.513\\
   &   & (3.91)  & (2.91)& \\
 Li$^+$  & F$_2$  & 7.783 &7.805 &7.807\\
   &   & (0.31)  & (0.03)& \\
 Li$^+$  & CO$_2$  & 9.387 &9.403  &9.396\\
   &   & (0.10)  & (-0.07)& \\
 Na$^+$  & H$_2$O  & 6.512  & 6.521 &6.527\\
   &   & (0.23)  & (0.07)& \\
 H$_3$O$^+$  & Ar & 2.681  &2.683 & 2.707\\
   &   & (0.96)  & (0.89)& \\
 NH$_4^+$  & Ar & 1.661  & 1.665&1.816\\
   &   & (8.52)  & (8.33)& \\
 Be$^2+$  & He  & 12.475  & 12.579 &12.986\\
   &   & (3.94)  & (3.14)& \\
 Be$^2+$  & H$_2$O  & 12.066  &12.336 & 13.569\\
   &   & (11.08)  & (9.09)& \\
 Mg$^2+$  & He  & 17.302  &17.313& 17.338\\
   &   & (0.20)  & (0.14)& \\
 Mg$^2+$  & H$_2$O  & 19.441  &19.511  & 19.751\\
   &   & (1.57)  & (1.22)& \\
\hline
\hline
\end{tabular}
\label{dipolemoments_charged.tab}
 \\
$^a$ for charged systems the dipole moment is calculated for
the cation at the origin.
\end{table}

\newpage

\begin{table}
\centering
\caption{
The effect of adding non-decomposable contribution to the embedding
potential
on the energy on the lowest unoccupied orbital localised in the
environment (LUEO). 
For each complex, the used electron density of the environment
($\rho_B$)
 is the ground-state Kohn-Sham (LDA) electron
density of the isolated subsystem $B$.
}
\begin{tabular}{ccrrr}
\hline
subsystem $B$  & subsystem $A$&$\epsilon_{LUEO}$ 
&$\epsilon_{LUEO}$
&$\Delta\epsilon_{LUEO}$\\
 & &$\tilde{v}_s=\tilde{v}^{TF}$
&$\tilde{v}_s=\tilde{v}^{NDSD}$
&\\
\hline
\hline
Li$^+$  & F$^-$  & -2.249  & -2.160 & 0.089 \\
Li$^+$  & Cl$^-$  & -2.669  &-2.568 & 0.101\\
Li$^+$  & Br$^-$  & -2.741  &-2.635  & 0.106\\
Na$^+$  & F$^-$  & -1.908 & -1.874& 0.034 \\
Na$^+$  & Cl$^-$  & -2.202 & -2.164& 0.038\\
Na$^+$  & Br$^-$  & -2.253  &-2.214 &0.039\\
Be$^{2+}$  & O$^{2-}$  &-4.922  &-4.913 & 0.009\\
Mg$^{2+}$  & O$^{2-}$  & -4.634  &-4.609 &0.025\\
\hline
\hline
\end{tabular}
\label{lueo_shifts_1}
\end{table}

\newpage

\begin{table}[h]
\centering
\caption{
The effect of adding non-decomposable contribution to the embedding
potential
on the energy on the lowest unoccupied orbital localised in the
environment (LUEO). 
For each complex, the used electron density of the environment
($\rho_B$)
 is the ground-state Kohn-Sham (LDA) electron
density of the isolated subsystem $B$.
}
\begin{tabular}{ccrrr}
\hline
subsystem $B$  & subsystem $A$&$\epsilon_{LUEO}$ 
&$\epsilon_{LUEO}$
&$\Delta\epsilon_{LUEO}$\\
 & &$\tilde{v}_s=\tilde{v}^{TF}$
&$\tilde{v}_s=\tilde{v}^{NDSD}$
&\\
\hline
\hline
Li$^+$  & H$_2$O  & -7.533 & -7.332& 0.201\\
Li$^+$  & F$_2$  & -8.501 & -8.223& 0.279\\
Li$^+$  & CO$_2$  & -9.069 & -8.810 & 0.259\\
Na$^+$  & H$_2$O  &-6.295  & -6.240&0.055 \\
Be$^2+$  & He  & -26.681  & -25.797& 0.884\\
Mg$^2+$  & He  & -18.376  &-18.078  & 0.299 \\
Be$^2+$  & H$_2$O  & -21.737  &-21.172 &0.565\\
Mg$^2+$  & H$_2$O  & -16.570  &-16.327 &0.242\\ 
\hline 
\hline
\end{tabular}
\label{lueo_shifts_3}
\end{table}

\newpage

\begin{table}
\centering
\caption{
The effect of adding non-decomposable contribution to the embedding
potential
on the energy on the highest occupied embedded orbital (HOEO). 
For each complex, the used electron density of the environment
($\rho_B$)
 is the ground-state Kohn-Sham (LDA) electron
density of the isolated subsystem $B$.
}
\begin{tabular}{ccccr}
\hline
subsystem $B$  & subsystem $A$&$\epsilon_{HOEO}$ 
&$\epsilon_{HOEO}$
&$\Delta\epsilon_{HOEO}$\\
 & &$\tilde{v}_s=\tilde{v}^{TF}$
&$\tilde{v}_s=\tilde{v}^{NDSD}$
&\\
\hline
\hline
 Li$^+$  & F$^-$  & -6.716  & -6.635 & 0.081 \\
 Li$^+$  & Cl$^-$  & -6.399  &-6.339 & 0.060\\
 Li$^+$  & Br$^-$  & -6.122  & -6.068 & 0.054\\
 Na$^+$  & F$^-$  & -5.223 & -5.198& 0.025 \\
 Na$^+$  & Cl$^-$  &-5.397  & -5.379& 0.018\\
 Na$^+$  & Br$^-$  & -5.277  &-5.261 &0.016\\
 Be$^{2+}$  & O$^{2-}$  &-6.724  &-6.690 & 0.033\\
 Mg$^{2+}$  & O$^{2-}$  & -5.135  &-5.108 &0.027\\ 
 H$_2$O$^a$  & H$_2$O  & -6.799  &-6.799 & 0.000\\
 HF$^a$  & HF  & -9.450  &-9.450  & 0.000\\
 H$_2$O$^b$  & H$_2$O  & -8.010  &-8.011 & -0.001\\
 HF$^b$  & HF  & -10.896  &-10.896  & 0.000\\
\hline
\hline
\end{tabular}
\label{hoeo_shifts_1}
 \\
$^a$ acceptor of hydrogen bond\\
$^b$ donor of hydrogen bond\\
\end{table}

\newpage 

\begin{table}[h]
\centering
\caption{
The effect of adding non-decomposable contribution to the embedding
potential
on the energy on the highest occupied embedded orbital (HOEO). 
For each complex, the used electron density of the environment
($\rho_B$)
 is the ground-state Kohn-Sham (LDA) electron
density of the isolated subsystem $B$.
}
\begin{tabular}{ccccr}
\hline
subsystem $B$  & subsystem $A$&$\epsilon_{HOEO}$ 
&$\epsilon_{HOEO}$
&$\Delta\epsilon_{HOEO}$\\
 & &$\tilde{v}_s=\tilde{v}^{TF}$
&$\tilde{v}_s=\tilde{v}^{NDSD}$
&\\
\hline
\hline
 Li$^+$  & H$_2$O  & -14.530 &-14.492& 0.037\\
 Li$^+$  & F$_2$  & -16.507 & -16.493& 0.014\\
 Li$^+$  & CO$_2$  & -14.749 & -14.742 &0.007\\
 Na$^+$  & H$_2$O  &-13.284  & -13.278 &0.006 \\
 Be$^2+$  & He  &-37.558  &-37.306 & 0.252\\
 Mg$^2+$  & He  & -31.410  &-31.386  & 0.024 \\
 Be$^2+$  & H$_2$O  & -25.208  &-25.001 &0.207\\
 Mg$^2+$  & H$_2$O  & -21.178  &-21.131 &0.047\\
\hline
\hline
\end{tabular}
\label{hoeo_shifts_3}
\end{table}

\newpage
\begin{table}[h]
\centering
\caption{
Total dipole moment ($\mu$ in Debye) obtained from the {\it freeze-and-thaw}
calculations using 
different approximations $v_t[\rho_A,\rho_B](\vec{r})$. 
The target Kohn-Sham results are given in
Tables~\ref{dipolemoments.tab}
and ~\ref{dipolemoments_charged.tab}.
Only dipole
moment components along the principal axis connecting the subsystems are
given.
Relative percentage errors of the dipole moments derived from
subsystem based calculations ($\frac{\Delta \mu^{subsystem}-\Delta
  \mu^{Kohn-Sham}}{\Delta \mu^{Kohn-Sham}}100\%$)
are given in parentheses.
}
$\;$\\
\begin{tabular}{cc|rr||rr}
\hline
\hline
& & & & & \\
\multicolumn{2}{c|}{system}&$v^{GEA2}_t$& $v^{GEA2}_t+f\cdot
v_t^{limit}$  & $v^{GGA97}_t$&$v^{GGA97}_t+f\cdot v_t^{limit}$\\
& & & & & \\
\hline
Li$^+$  & Cl$^-$  & 5.365  &5.519 &5.671 & 5.829\\
  &   & (21.4)  & (19.2)& (16.9) & (14.6) \\
Li$^+$  & H$_2$O$^a$  & -$^b$  &5.188&5.252 & 5.303\\
  &   & (-)  & (5.9)& (4.7) & (3.8)\\
Na$^+$  & Cl$^-$  & -$^b$  &8.025 &8.346 &8.400 \\
  &   & (-)  & (6.9)& (3.0) & (2.4)\\
Na$^+$  & H$_2$O$^a$ & 6.378  &6.388  &6.459 &6.470 \\
  &   & (2.3) & (2.1)& (1.0) & (0.9)\\
Be$^{2+}$  & O$^{2-}$  & 3.962  &4.073 & 4.128 & 4.245\\
  &   & (35.3)  & (33.4)& (32.5)& (30.6) \\
HF  & HF  & 2.994  &2.994 & 2.995& 2.995\\
  &   & (3.1)  & (3.1)& (3.1) & (3.1) \\ 
\hline
\hline
\end{tabular}\label{other.decomposable}
\label{other_decomposable}
\\
$^a$ for charged systems the dipole moment is calculated for
the cation at the origin.
$^b$ no convergence.
\end{table}

\newpage

\newpage

\newpage
\end{document}